\documentclass{article}

\usepackage{PRIMEarxiv}
\usepackage[utf8]{inputenc}
\usepackage[T1]{fontenc}
\usepackage{amsmath,amssymb,amsthm}
\usepackage[numbers,square,sort&compress]{natbib}
\usepackage{hyperref}
\usepackage{url}
\usepackage{booktabs}
\usepackage{nicefrac}
\usepackage{microtype}
\usepackage{graphicx}
\usepackage{epstopdf}
\usepackage{algorithmic}
\usepackage{quantikz}
\usepackage{cancel}
\usepackage[capitalize,nameinlink]{cleveref}
\usetikzlibrary{decorations.pathreplacing}
\DeclareGraphicsExtensions{.pdf,.png,.jpg,.eps}

\renewcommand{\keywordname}{{\bfseries\emph{Key words}}}

\numberwithin{equation}{section}
\newcommand{\funding}[1]{\protect\\\textbf{Funding:} #1}
\title{Rotation Collision based Quantum Lattice Boltzmann Methods\thanks{Submitted to the editors DATE.
\funding{This work was financially supported by the National Natural Science Foundation of China (Grant Nos. 12501599 and 123B2018), the Postdoctoral Fellowship Program of CPSF
(Grant No. GZB20250714), China Postdoctoral Science Foundation (Grant No. 2025M773077), the Open Research Fund of State Key Laboratory of Mesoscience and Process Engineering (Grant No.
MESO-25-D04) and the Interdisciplinary Research Program of Hust (Grant No. 2024JCYJ001).
}}}
\author{%
  \parbox{0.94\textwidth}{%
    \centering\normalfont
    Kangyang Zeng\textsuperscript{1},
    Changshen Huang\textsuperscript{1},
    Xi Liu\textsuperscript{1},
    Xiaodong Niu\textsuperscript{2},
    and Zhenhua Chai\textsuperscript{1,3,4,5}\thanks{Corresponding author.}\\[2pt]
    \small
    \textsuperscript{1}School of Mathematics and Statistics, Huazhong University of Science and Technology, Wuhan 430074, China\\[2pt]
    \textsuperscript{2}College of Engineering, Shantou University, Shantou 515063, China\\[2pt]
    \textsuperscript{3}Institute of Interdisciplinary Research for Mathematics and Applied Science, Huazhong University of Science and Technology, Wuhan 430074, China\\[2pt]
    \textsuperscript{4}Hubei Key Laboratory of Engineering Modeling and Scientific Computing, Huazhong University of Science and Technology, Wuhan 430074, China\\[2pt]
    \textsuperscript{5}The State Key Laboratory of Intelligent Manufacturing Equipment and Technology, Huazhong University of Science and Technology, Wuhan 430074, China\\[2pt]
    \texttt{kyzeng@hust.edu.cn}, \texttt{hcs@hust.edu.cn}, \texttt{aubrey\_xixi@126.com}\\[2pt]
    \texttt{xdniu@stu.edu.cn}, \texttt{hustczh@hust.edu.cn}
  }%
}

\hypersetup{
  pdftitle={Rotation Collision based Quantum Lattice Boltzmann Methods},
  pdfauthor={K. Zeng, C. Huang, X. Liu, Z. Chai, and X. Niu},
  colorlinks=true,
  linkcolor=black,
  citecolor=black,
  filecolor=black,
  urlcolor=black
}

\begin{document}

\maketitle

\begin{abstract}
In the existing quantum algorithms for the lattice Boltzmann method, resolving the unitarity problem of the BGK collision operator from first principles remains a fundamental challenge, and a systematically constructed 
unitary alternative that reproduces the BGK relaxation structure to leading order has not yet been established in the literature. In this work, we propose a novel quantum lattice Boltzmann method (QLBM) based on the rotation collision operator, in which the non-unitary BGK relaxation is replaced by a unitary rotation in the amplitude space of the distribution function, and the corresponding gate-level quantum circuits are also constructed. Specifically, the rotation collision operator is developed for both the D2Q5 model of the convection-diffusion equation and the D2Q9 model of the incompressible Navier--Stokes equations. It is worth noting that, for the D2Q5 model, the equilibrium-state preparation circuit can be precompiled and reused, and the resulting unitary blocks can be sequentially composed once the state-dependent collision parameters are specified. Finally, some numerical experiments are performed to validate the developed QLBM, demonstrating its accuracy and effectiveness.
\end{abstract}


\section{Introduction}
Over the past three decades, the lattice Boltzmann method (LBM) has 
emerged as a competitive numerical tool in computational fluid 
dynamics, and has been widely applied to a broad range of fluid flow 
and heat/mass transfer problems \cite{chen1998lattice,succi2001lattice}. As a mesoscopic approach, the LBM not only can recover the macroscopic governing equations in the hydrodynamic limit, but also retains a collision–streaming structure that is well suited for parallel 
computing \cite{kruger2017lattice}.

Recent advances in the theory and hardware of quantum computing \cite{Preskill2018quantumcomputingin, arute2019quantum} have motivated researchers to explore quantum algorithms for 
scientific computing, and computational fluid dynamics (CFD) has emerged as one of the most promising application areas \cite{cao2013quantum, harrow2009quantum}. In particular, 
several quantum CFD methods have been proposed for the simulation of fluid flows governed by the Euler or Navier--Stokes equations \cite{gaitan2020finding}, exploiting the potential quantum speedup in solving large-scale linear systems \cite{harrow2009quantum}. Among these approaches, the LBM provides a natural bridge to quantum computing, because its collision-streaming structure maps to quantum circuits naturally \cite{mezzacapo2015quantum, yepez2001quantum}. For this reason, several quantum lattice Boltzmann methods (QLBMs) have been developed for the convection-diffusion equation and the Navier--Stokes equations \cite{ljubomir2022quantum, wawrzyniak2025quantum}. However, in such implementations, there are two fundamental problems that need to be addressed. The first one is the unitarity issue, i.e., the quantum evolution is required to be unitary, whereas the BGK collision operator in LBM is dissipative and non-unitary. The second is the nonlinearity issue, i.e., the equilibrium distribution function in the D2Q9 model contains some quadratic terms of the fluid velocity, which cannot be directly encoded in quantum circuits without violating the no-cloning theorem \cite{wootters1982single}. To overcome these two problems, several strategies have been developed in some previous works. For instance, Mezzacapo et al. \cite{mezzacapo2015quantum} first decomposed the non-unitary collision operator into a linear combination of unitary operators and then implemented it probabilistically with the help of an ancilla qubit, while the success probability of this 
approach decreases exponentially with the system size. Unlike this method, Todorova and Steijl \cite{todorova2020quantum} avoided the collision term by adopting a collisionless model, which, however, fundamentally restricts the physical scope of the method. Later, Zeng \cite{zeng2025quantum,zeng2026quantum} proposed a linearized collision operator within the QLBM framework, but the linearization inevitably introduces 
approximation error that deviates from the original physical model. Xiao et al. \cite{xiao2026interpolation,xiao2026stable} further extended QLBMs to incompressible flows on nonuniform meshes and in domains with curved boundaries, and improved numerical stability at high Reynolds numbers via fractional-step formulations. Kumar and Frankel \cite{kumar2025quantum} represented the collision operator as a unitary matrix via 
singular value decomposition, achieving an exact unitary encoding. However, the resulting matrix depends on the current quantum state and must be recomputed at each time step, with gate complexity growing exponentially in the number of qubits. In addition, Wang et al.~\cite{wang2025quantum} proposed a novel ensemble method based on the lattice gas automaton, which provides an alternative perspective for the quantum implementation of the LBM. Moreover, it is worth mentioning that such a linearized operator no longer preserves the genuine BGK structure at the level of quantum encoding. Nevertheless, to the best of our knowledge, none of these approaches that can be used to solve the unitarity problem of the collision operator from the first principles.

In this work, we develop a quantum lattice Boltzmann framework based on a rotation collision operator used to address the unitarity issue of the BGK relaxation. The proposed framework provides a unitary formulation of the collision process in the amplitude space and corresponding quantum circuit implementations for both the D2Q5 model of the convection--diffusion equation and the D2Q9 model of the incompressible Navier--Stokes equations. 

The rest of the paper is organized as follows. In Section 2, the rotation collision operator is introduced, followed by its quantum circuit implementations for the D2Q5 and D2Q9 models. Numerical results are presented in Section 4, and some conclusions are summarized in Section 5. Finally, the gate-level circuit constructions are provided in the Appendices.

\section{Rotation Collision Operator}
\label{sec:rotation}

\subsection{Lattice Boltzmann Method}
In the lattice Boltzmann method with the BGK collision operator, also called lattice BGK method, the distribution function
$\{f_i\}_{i=0}^{Q-1}$ at a single lattice site is updated according to
the lattice Boltzmann equation (LBE) \cite{chen1998lattice,qian1992lattice},
\begin{equation}\label{eq:lbe}
  f_i(\boldsymbol{x}+\boldsymbol{e}_i\delta t,\, t+\delta t)
  = f_i(\boldsymbol{x},t)
    - \frac{1}{\tau}\bigl(f_i(\boldsymbol{x},t)
    - f_i^{eq}(\boldsymbol{x},t)\bigr),
\end{equation}
where $Q$ is the number of discrete velocity directions,
$\boldsymbol{e}_i$ is the $i$-th discrete velocity, $\delta t$ is
the time step, $\tau$ is the relaxation time, and $f_i^{eq}$ is the
equilibrium distribution function.
The macroscopic density and velocity are detemined by the moments of the
distribution function,
\begin{equation}\label{eq:macro}
  \rho = \sum_i f_i, \qquad \rho\boldsymbol{u} = \sum_i f_i\boldsymbol{e}_i.
\end{equation}
The equilibrium distribution function satisfies the following conservation
conditions \cite{kruger2017lattice},
\begin{equation}\label{eq:conservation}
  \sum_i f_i^{eq} = \rho, \qquad
  \sum_i f_i^{eq}\boldsymbol{e}_i = \rho\boldsymbol{u}.
\end{equation}
Based on \cref{eq:lbe}, the collision step in Eq. \cref{eq:lbe} can be written as
\begin{equation}\label{eq:bgk}
  f_i^{*} = f_i - \frac{1}{\tau}(f_i - f_i^{eq})
           = \Bigl(1-\frac{1}{\tau}\Bigr)f_i + \frac{1}{\tau}f_i^{eq},
\end{equation}
which is a linear interpolation between $f_i$ and $f_i^{eq}$.
Thus, the deviation from equilibrium state is reduced by a factor of
$(1-1/\tau)$ at each collision step,
\begin{equation}\label{eq:bgk-relax}
  f_i^{*} - f_i^{eq} = \Bigl(1-\frac{1}{\tau}\Bigr)(f_i - f_i^{eq}).
\end{equation}

\subsection{Non-Unitarity of the BGK Collision Operator}
In quantum computing, each admissible operation must be
implemented as a unitary operator \cite{nielsen2010quantum}.
The BGK collision operator, however, presents two fundamental
obstacles that prevent its direct implementation as a quantum gate.

The first one is the nonlinearity in amplitude space.
The natural encoding of the distribution function into a quantum
state is given by the amplitudes $\psi_i = \sqrt{f_i}$
\cite{mezzacapo2015quantum}, that is,
$|\psi\rangle = \sum_i \sqrt{f_i}\,|i\rangle$.
Under this encoding, the BGK collision operator takes the form
\begin{equation}\label{eq:bgk-amp}
  \sqrt{f_i^{*}} = \sqrt{\Bigl(1-\frac{1}{\tau}\Bigr)f_i
                   + \frac{1}{\tau}f_i^{eq}},
\end{equation}
and due to the fact $\sqrt{a+b} \neq \sqrt{a}+\sqrt{b}$ in general, this
is not equal to $(1-1/\tau)\sqrt{f_i} + (1/\tau)\sqrt{f_i^{eq}}$.
This means that the BGK collision operator is nonlinear in amplitude
space and cannot be represented as a linear operator on quantum
state.

The second obstacle is the non-unitarity of linear
interpolation. Even we consider the corresponding linear
interpolation directly in amplitude space, such an operation is
not unitary. Specifically, for any
$|\psi\rangle \neq |\psi_{eq}\rangle$, we have
\begin{equation}\label{eq:norm-violation}
  \Bigl\|\Bigl(1-\frac{1}{\tau}\Bigr)|\psi\rangle
        + \frac{1}{\tau}|\psi_{eq}\rangle\Bigr\|
        \neq \||\psi\rangle\|,
\end{equation}
which indicates that the linear interpolation does not preserve the norm of
the state vector. This is in conflict with the requirement that
any quantum operation must satisfy
$\|U|\psi\rangle\| = \||\psi\rangle\|$.
From the above analysis, it can be seen that the BGK collision
operator cannot be directly realized as a quantum gate, and a
unitary alternative is therefore required.

\subsection{Amplitude Encoding}

Given a single-site distribution function $\{f_i\}_{i=0}^{Q-1}$,
the corresponding normalized quantum state \cite{schuld2019quantum} is defined by
\begin{equation}\label{eq:encoding}
  |\psi\rangle = \frac{1}{\sqrt{\rho}}\sum_{i=0}^{Q-1}\sqrt{f_i}\,|i\rangle,
  \qquad \langle\psi|\psi\rangle = 1,
\end{equation}
from which the distribution function can be recovered by
$f_i = \rho\,|\langle i|\psi\rangle|^2$.
Similarly, the equilibrium distribution function is encoded as
\begin{equation}\label{eq:eq-encoding}
  |\psi_{\mathrm{eq}}\rangle = \frac{1}{\sqrt{\rho}}
  \sum_{i=0}^{Q-1}\sqrt{f_i^{\mathrm{eq}}}\,|i\rangle,
\end{equation}
and it can be verified that $|\psi_{\mathrm{eq}}\rangle$ is also 
a normalized state by the conservation condition Eq.~\cref{eq:conservation}.

\subsection{Construction of the Rotation Collision Operator}
\label{sec:construction}
Since the amplitudes $\sqrt{f_i}$ and $\sqrt{f_i^{\mathrm{eq}}}$ are
real and non-negative, the inner product
$\langle\psi|\psi_{\mathrm{eq}}\rangle$ is a real number in
$[0,1]$, and $|\psi\rangle$ and $|\psi_{\mathrm{eq}}\rangle$ are
both unit vectors in the Hilbert space. We define the angle between
them as
\begin{equation}\label{eq:angle}
  \phi = \arccos\langle\psi|\psi_{\mathrm{eq}}\rangle \in [0,\pi/2].
\end{equation}
Based on \cref{eq:bgk-relax}, the BGK collision reduces the distance
between $f$ and $f^{\mathrm{eq}}$ by a factor of $(1-1/\tau)$ at
each step. In the Hilbert space representation, this deviation is characterized by the angle $\phi$ between $|\psi\rangle$ and
$|\psi_{\mathrm{eq}}\rangle$, the unitary analogue of the BGK collision is therefore expected to contract the angle $\phi$ by the same factor corresponding to a rotation of
$|\psi\rangle$ toward $|\psi_{\mathrm{eq}}\rangle$ such that the angle
decreases from $\phi$ to
\begin{equation}\label{eq:theta}
  \phi\Bigl(1-\frac{1}{\tau}\Bigr) = \phi - \theta,
  \qquad \theta = \frac{\phi}{\tau}.
\end{equation}
If $\phi = 0$, the state $|\psi\rangle$ already coincides with
$|\psi_{\mathrm{eq}}\rangle$, $\theta = 0$, and
$U_{\mathrm{col}} = I$ trivially. In what follows assume $\phi\neq 0$, and accordingly define the unit vector in the
plane spanned by $|\psi\rangle$ and $|\psi_{\mathrm{eq}}\rangle$
that is orthogonal to $|\psi\rangle$ and in the direction of
$|\psi_{\mathrm{eq}}\rangle$ as
\begin{equation}\label{eq:perp}
  |\psi_\perp\rangle =
  \frac{|\psi_{\mathrm{eq}}\rangle - \cos\phi\,|\psi\rangle}{\sin\phi}.
\end{equation}
Substituting \cref{eq:angle} gives
$\langle\psi|\psi_\perp\rangle
= (\langle\psi|\psi_{\mathrm{eq}}\rangle - \cos\phi)/\sin\phi = 0$
and $\langle\psi_\perp|\psi_\perp\rangle = 1$, thus
$\{|\psi\rangle, |\psi_\perp\rangle\}$ forms an orthonormal basis
for the two-dimensional subspace
$V = \mathrm{span}\{|\psi\rangle, |\psi_{\mathrm{eq}}\rangle\}$.
The rotation collision operator is then defined as the rotation by
angle $\theta$ within $V$ and the identity on its orthogonal
complement,
\begin{equation}\label{eq:ucol}
  U_{\mathrm{col}} = I + (\cos\theta-1)
            \bigl(|\psi\rangle\langle\psi|
            + |\psi_\perp\rangle\langle\psi_\perp|\bigr)
            + \sin\theta\bigl(|\psi_\perp\rangle\langle\psi|
            - |\psi\rangle\langle\psi_\perp|\bigr),
\end{equation}
and the post-collision quantum state is given by
\begin{equation}\label{eq:psi-prime}
  |\psi'\rangle = U_{\mathrm{col}}|\psi\rangle
               = \cos\theta\,|\psi\rangle
               + \sin\theta\,|\psi_\perp\rangle,
\end{equation}
from which the post-collision distribution function is recovered by
$f_i^{*} = \rho\,|\langle i|\psi'\rangle|^2$.

It is clear that \cref{eq:ucol} defines $U_{\mathrm{col}}$ in terms of the
explicit vector $|\psi_\perp\rangle$, which does not need to
be prepared as a separate quantum state at the circuit level: an
equivalent realization of $U_{\mathrm{col}}$ can be assembled from
the state-preparation circuits of $|\psi\rangle$ and
$|\psi_{\mathrm{eq}}\rangle$ alone, using a sequence of generalized
reflections; the details of this circuit construction are given in Appendix B.

The rotation collision operator $U_{\mathrm{col}}$ defined in \cref{eq:ucol} is unitary, i.e.,
$U_{\mathrm{col}}^\dagger U_{\mathrm{col}} = I$. This follows from
the fact that the restriction of $U_{\mathrm{col}}$ to the subspace
$V$ is the rotation matrix
$R=\bigl(\begin{smallmatrix}\cos\theta & -\sin\theta \\
\sin\theta & \cos\theta\end{smallmatrix}\bigr)$, which satisfies
$R^\top R = I$, while $U_{\mathrm{col}}$ acts as the identity on the
orthogonal complement of $V$.

As a direct consequence of the unitarity of $U_{\mathrm{col}}$, the
rotation collision operator conserves mass exactly. Indeed, one can obtain
$\sum_i f_i^{*} = \rho\sum_i|\langle i|\psi'\rangle|^2 = \rho$ since $\||\psi'\rangle\| = \||\psi\rangle\| = 1$ .

In addition, we can show that the inner product of the post-collision state $|\psi'\rangle$ with
$|\psi_{\mathrm{eq}}\rangle$ also satisfies the following relation,
\begin{equation}\label{eq:angle-after}
  \langle\psi'|\psi_{\mathrm{eq}}\rangle
  = \cos\theta\cos\phi + \sin\theta\sin\phi
  = \cos(\phi-\theta),
\end{equation}
so the post-collision angle is $\phi-\theta = \phi(1-1/\tau)$,
which agrees exactly with the relaxation rate $(1-1/\tau)$ of the
BGK collision operator in \cref{eq:bgk-relax}.

The rotation collision operator $U_{\mathrm{col}}$ is an exact unitary
transformation in the Hilbert space. The difference from the classical
BGK collision arises when the amplitude-space evolution is mapped back
to the distribution-function space through
\begin{equation}\label{eq:recovery-rot}
  f_i^{\mathrm{rot}}
  =
  \rho\,|\langle i|\psi'\rangle|^2 .
\end{equation}
We show below that, in the continuum limit $\delta t\to0$, the
rotation collision agrees with the BGK collision up to first order
in $\delta t$.

Consider the evolution equation \cref{eq:lbe}, and introduce $D_i=\partial_t+\boldsymbol{e}_i\cdot\nabla$ , we can obtain the following equation with the help of Taylor expansion
\begin{equation}\label{eq:lbe-taylor}
  \delta t\,D_i f_i
  +
  \frac{\delta t^2}{2}D_i^2f_i
  +
  O(\delta t^3)
  =
  -\frac{1}{\tau}
  \left(f_i-f_i^{\mathrm{eq}}\right).
\end{equation}
From above equation one can derive \cite{chai2020multiple}
\begin{equation}\label{eq:noneq-dt}
  f_i-f_i^{\mathrm{eq}}=O(\delta t).
\end{equation}
Under the condition of non-negative equilibrium/distribution functions, and let
$a_i=\sqrt{f_i/\rho}$ and
$a_i^{\mathrm{eq}}=\sqrt{f_i^{\mathrm{eq}}/\rho}$ denote the
normalized amplitudes of the distribution and equilibrium states,
respectively, the following equation can be obtained from 
\cref{eq:noneq-dt} 
\begin{equation}
  a_i-a_i^{\mathrm{eq}}=O(\delta t).
\end{equation}
Since the two quantum states are normalized,
\begin{equation}\label{eq:cos-phi}
  \cos\phi
  =
  \langle\psi|\psi_{\mathrm{eq}}\rangle
  =
  1-\frac{1}{2}
  \sum_i
  \left(a_i-a_i^{\mathrm{eq}}\right)^2 ,
\end{equation}
we have
\begin{equation}\label{eq:phi-dt}
  \phi=O(\delta t).
\end{equation}

Substituting $|\psi_\perp\rangle$ into the post-collision state yields
\begin{equation}\label{eq:psi-prime-exact}
  |\psi'\rangle
  =
  \left(
  \cos\theta
  -
  \frac{\sin\theta\cos\phi}{\sin\phi}
  \right)|\psi\rangle
  +
  \frac{\sin\theta}{\sin\phi}
  |\psi_{\mathrm{eq}}\rangle .
\end{equation}
Consider $\theta=\phi/\tau$ and $\phi=O(\delta t)$, we have
\begin{equation}
  \frac{\sin\theta}{\sin\phi}
  =
  \frac{1}{\tau}
  +
  O(\delta t^2),
  \qquad
  \cos\theta
  -
  \frac{\sin\theta\cos\phi}{\sin\phi}
  =
  1-\frac{1}{\tau}
  +
  O(\delta t^2),
\end{equation}
which yields
\begin{equation}\label{eq:linear-interp}
  |\psi'\rangle
  =
  \left(1-\frac{1}{\tau}\right)|\psi\rangle
  +
  \frac{1}{\tau}|\psi_{\mathrm{eq}}\rangle
  +
  O(\delta t^2).
\end{equation}

Taking the $i$th component and recovering the distribution function, together
with
\begin{equation}
  \sqrt{f_i f_i^{\mathrm{eq}}}
  =
  \frac{f_i+f_i^{\mathrm{eq}}}{2}
  +
  O(\delta t^2),
\end{equation}
gives
\begin{equation}\label{eq:f-rot-result}
  f_i^{\mathrm{rot}}
  =
  f_i
  +
  \frac{f_i^{\mathrm{eq}}-f_i}{\tau}
  +
  O(\delta t^2).
\end{equation}
On the other hand, the post-collision distribution function is given by
\begin{equation}
  f_i^{\mathrm{BGK}}
  =
  f_i
  +
  \frac{f_i^{\mathrm{eq}}-f_i}{\tau},
\end{equation}
thus we have
\begin{equation}\label{eq:rot-bgk-order}
  f_i^{\mathrm{rot}}
  -
  f_i^{\mathrm{BGK}}
  =
  O(\delta t^2).
\end{equation}
From above discussion, one can find that compared to the BGK collision, the rotation collision only introduces a discrepancy  with a second order of time step, which is consistent with truncation error of the standard LBM.

In particular, if we consider the case of $\tau=1$, one can obtain $\theta=\phi$ and
$|\psi'\rangle=|\psi_{\mathrm{eq}}\rangle$ exactly. Consequently,
\begin{equation}
  f_i^{\mathrm{rot}}
  =
  f_i^{\mathrm{eq}}
  =
  f_i^{\mathrm{BGK}}
\end{equation}

\subsection{Quantum Circuit Implementation}
\label{sec:circuit-implementation}
For the D2Q9 model, the equilibrium state $|\psi_{\mathrm{eq}}\rangle$
depends on the density and velocity moments of the current state
$|\psi\rangle$. As a result, the collision operator
$U_{\mathrm{col}}$ is a $16\times 16$ state-dependent unitary that
must be updated at each time step according to the macroscopic
moments and applied to the four-qubit register encoding
$|\psi\rangle$.

In contrast, the equilibrium state
$|\psi_{\mathrm{eq}}\rangle$ in the D2Q5 model reduces to a fixed vector independent
of the current state (see \crefname{subsection}{Section}{Sections} \cref{sec:d2q5-linear}). Consequently,
$U_{\mathrm{col}}$ becomes a fixed unitary acting on the
three-qubit register, and can be precomputed once at the circuit
construction stage.

In this case, $U_{\mathrm{col}}$ admits a direct gate-level
synthesis using the Shende--Bullock--Markov state-preparation scheme
\cite{shende2005synthesis} together with multi-controlled phase
gates \cite{nielsen2010quantum}, without requiring dynamic
recompilation. The resulting gate-level circuits for both models are
provided in Appendices B and C.

\section{Quantum Lattice Boltzmann Method and Gate-Concatenability}
\label{sec:qlbm}
\subsection{Full Quantum LBM Evolution Structure}
At each time step, the classical LBM updates the distribution function through two
successive steps: a collision step followed by
a streaming step. In the quantum framework, each step corresponds
to a unitary operator acting on the joint state
\begin{equation}\label{eq:joint-state}
  |\Psi\rangle \in \mathcal{H}_{\mathrm{dir}}\otimes\mathcal{H}_{\mathrm{grid}}.
\end{equation}
The Hilbert space can be decomposed as
$\mathcal{H} = \mathcal{H}_{\mathrm{dir}} \otimes \mathcal{H}_{\mathrm{grid}}$,
where $\mathcal{H}_{\mathrm{dir}}$ encodes the $Q$ discrete velocity
directions at each lattice site, as defined in Eq. \cref{eq:encoding},
and $\mathcal{H}_{\mathrm{grid}}$ encodes the lattice coordinates.
The single-step quantum evolution operator is then given by
\begin{equation}\label{eq:ustep}
  U_{\mathrm{step}} = U_{\mathrm{stream}} \cdot U_{\mathrm{col}},
\end{equation}
where $U_{\mathrm{col}}$ is the rotation collision operator acting on $\mathcal{H}_{\mathrm{dir}}$ at
each site, and $U_{\mathrm{stream}}$ denotes the streaming operator defined below, which acts jointly on $\mathcal{H}_{\mathrm{dir}}\otimes\mathcal{H}_{\mathrm{grid}}$. The operator ordering in \cref{eq:ustep} reflects that
$U_{\mathrm{col}}$ is applied prior to $U_{\mathrm{stream}}$,
consistent with the classical collide-then-stream update.
The evolution over $T$ steps is then given by
\begin{equation}\label{eq:T-step}
  |\Psi(T)\rangle = U_{\mathrm{step}}^T|\Psi(0)\rangle.
\end{equation}

\subsection{Quantum Implementation of the Streaming Step}
The streaming step shifts the distribution function at one lattice
spacing along each discrete velocity direction by
\begin{equation}\label{eq:stream}
  f_i(\boldsymbol{x}+\boldsymbol{e}_i\delta t,\, t+\delta t)
  = f_i^*(\boldsymbol{x}, t).
\end{equation}
In the quantum framework, the streaming step corresponds to a
permutation operator $U_{\mathrm{stream}}$ acting jointly on the
direction and lattice-coordinate registers. We describe two
implementations that realize this permutation by different
circuit constructions.

\textbf{Scheme 1: Quantum walk \cite{childs2010relationship,portugal2013quantum}.}
The streaming step is represented as a quantum walk with the
discrete velocity direction serving as the coin degree of freedom.
Actually, the streaming operator can be decomposed as
\begin{equation}\label{eq:qwalk}
  U_{\mathrm{stream}} = S \cdot (C \otimes I_{\mathrm{grid}}),
\end{equation}
where $C$ acts on $\mathcal{H}_{\mathrm{dir}}$ as the coin operator, 
and $S$ is the conditional shift operator that translates the
lattice register by one site along each velocity direction. Note that $C$ is distinct from the collision operator $U_{\mathrm{col}}$
in \cref{eq:ustep} and serves only to route the shift $S$, not to
relax the distribution function toward its equilibrium state. Under periodic boundary
conditions, the conditional shift reduces to a modular addition,
which can be implemented efficiently with quantum adder circuits.

\textbf{Scheme 2: Quantum Fourier transform \cite{shakeel2020efficient,coppersmith2002approximate}.}
Under periodic boundary conditions, the spatial shift operator is
diagonal in the Fourier basis. The quantum Fourier transform
$\mathcal{F}$ diagonalizes the shift $T_{\boldsymbol{e}_i}$ as
\begin{equation}\label{eq:qft}
  \mathcal{F} \cdot T_{\boldsymbol{e}_i} \cdot \mathcal{F}^\dagger
  = \mathrm{diag}\!\left(e^{2\pi \mathrm{i} k e_i / N}\right)_{k=0}^{N-1},
\end{equation}
where $\mathrm{i}=\sqrt{-1}$ denotes the imaginary unit, written in
upright type to distinguish it from the direction index $i$ used
elsewhere. The streaming step is thus implemented in
three stages: the QFT, diagonal phase gates, and inverse QFT. The QFT circuit requires $O(n^2)$ gates with
$n=\log_2 N$, compared with $O(N\log N) = O(n\,2^n)$ for the
classical fast Fourier transform, giving rise to an exponential reduction
in gate count.

Both schemes can realize the same permutation operator
$U_{\mathrm{stream}}$ through different circuit constructions, and
both are applicable to the D2Q5 and D2Q9 models.

\subsection{Sequential Concatenation of QLBM Steps}
\label{sec:concat}

The implementation of QLBM consists of a
collision operation followed by a streaming operation. Since the
rotation collision operator is constructed from the instantaneous
state, we introduce a time-step index and write the
one-step evolution operator as
\begin{equation}\label{eq:ustep-t}
  U_{\mathrm{step},t}
  =
  U_{\mathrm{stream}} U_{\mathrm{col},t},
\end{equation}
where $U_{\mathrm{col},t}$ is the rotation collision operator
at time $t$, $U_{\mathrm{stream}}$ denotes the streaming
operator.

Accordingly, the evolution over $T$ time steps can be represented by
the ordered composition
\begin{equation}\label{eq:multistep-evolution}
  |\Psi(T)\rangle
  =
  U_{\mathrm{step},T-1}
  U_{\mathrm{step},T-2}
  \cdots
  U_{\mathrm{step},1}
  U_{\mathrm{step},0}
  |\Psi(0)\rangle ,
\end{equation}
or equivalently, 
\begin{equation}\label{eq:multistep-expanded}
  |\Psi(T)\rangle
  =
  \left[
  \prod_{t=T-1}^{0}
  U_{\mathrm{stream}} U_{\mathrm{col},t}
  \right]
  |\Psi(0)\rangle .
\end{equation}
We refer to this ordered composition of successive unitary
collision--streaming blocks as sequential circuit concatenation.
Unlike repeated application of a single time-independent operator,
this construction does not require the collision operators at
different time steps to be identical. It should be noted that when the complete one-step operator is independent of both the instantaneous state and the time step, the following special form can be recovered
\begin{equation}\label{eq:fixed-step-special}
  |\Psi(T)\rangle
  =
  U_{\mathrm{step}}^{T}|\Psi(0)\rangle
\end{equation}

Once the parameters defining the individual collision operators
are specified, the corresponding unitary blocks can be connected
sequentially without requiring a projective measurement solely for
the composition of adjacent quantum operations. The determination
of state-dependent collision parameters from an intermediate quantum
state is a separate issue from the unitary composition itself.

\subsection{Linearity of the D2Q5 Equilibrium Distribution for convection-diffusion equation}
\label{sec:d2q5-linear}

In this work, the D2Q5 model is used to solve the
convection--diffusion equation
\begin{equation}\label{eq:cde}
  \partial_t\phi + \nabla\cdot(\boldsymbol{v}\phi)
  = \kappa\nabla^2\phi,
\end{equation}
where $\boldsymbol{v}$ is an externally prescribed velocity field
and $\kappa=c_s^2(\tau-1/2)\delta t$ is the diffusion coefficient.
The equilibrium distribution function and macroscopic variable
$\phi$ are given by~\cite{wolf2004lattice}
\begin{equation}\label{eq:d2q5-eq}
  f_i^{\mathrm{eq}} = \lambda_i\phi,
  \qquad
  \lambda_i =
  w_i\!\left(
  1+\frac{\boldsymbol{e}_i\cdot\boldsymbol{v}}{c_s^2}
  \right),
  \qquad
  \phi=\sum_j f_j .
\end{equation}
The coefficients $\lambda_i$ depend only on the prescribed
parameters $\{w_i,\boldsymbol{e}_i,\boldsymbol{v}\}$ and are
independent of the instantaneous distribution $\{f_j\}$.
Consequently, $f_i^{\mathrm{eq}}$ is a linear function of the
distribution function. This structure leads to an important
simplification in the quantum representation: the normalized
equilibrium direction is independent of the instantaneous
distribution.

Indeed, the normalized equilibrium state can be written as
\begin{equation}\label{eq:eeq}
  |\hat e_{\mathrm{eq}}\rangle
  =
  \frac{
  \displaystyle\sum_i\sqrt{\lambda_i}\,|i\rangle
  }{
  \left(\displaystyle\sum_i\lambda_i\right)^{1/2}
  }.
\end{equation}
From \cref{eq:d2q5-eq} we have
$\sqrt{f_i^{\mathrm{eq}}}
=\sqrt{\lambda_i}\sqrt{\phi}$.
Since the common factor $\sqrt{\phi}$ cancels upon normalization,
the amplitude-encoded equilibrium state satisfies
\begin{equation}\label{eq:d2q5-eq-state}
  |\psi_{\mathrm{eq}}\rangle
  =
  |\hat e_{\mathrm{eq}}\rangle ,
\end{equation}
and is therefore independent of the instantaneous distribution.

\subsection{Rotation Collision and Sequential Evolution for the D2Q5 Model}
\label{sec:d2q5-rotation}

As shown in the \cref{sec:d2q5-linear}, the normalized
equilibrium state of the D2Q5 model is the fixed direction
$|\hat e_{\mathrm{eq}}\rangle$, independent of the instantaneous
distribution function. This property simplifies the implementation of the
rotation collision, although the complete collision operator
generally remains state dependent.

Let $|\psi_t\rangle$ denote the amplitude-encoded distribution
state at time $t$. The angle between the instantaneous state
and the fixed equilibrium direction is
\begin{equation}\label{eq:d2q5-phi-t}
  \phi_t
  =
  \arccos
  \langle\psi_t|\hat e_{\mathrm{eq}}\rangle .
\end{equation}
Following the rotation collision construction introduced in
\cref{sec:rotation}, the corresponding rotation angle is
\begin{equation}\label{eq:d2q5-theta-t}
  \theta_t
  =
  \frac{\phi_t}{\tau}.
\end{equation}
If $\phi_t\neq0$, the unit vector orthogonal to $|\psi_t\rangle$
within the two-dimensional rotation subspace
$\operatorname{span}\{|\psi_t\rangle,
|\hat e_{\mathrm{eq}}\rangle\}$
is given by
\begin{equation}\label{eq:d2q5-perp-t}
  |\psi_{\perp,t}\rangle
  =
  \frac{
    |\hat e_{\mathrm{eq}}\rangle
    -
    \cos\phi_t|\psi_t\rangle
  }{
    \sin\phi_t
  } .
\end{equation}
Accordingly, the rotation collision operator at time $t$
takes the form
\begin{align}
U_{\mathrm{col},t}
={}&
I
+
(\cos\theta_t-1)
\left(
|\psi_t\rangle\langle\psi_t|
+
|\psi_{\perp,t}\rangle
\langle\psi_{\perp,t}|
\right)
\nonumber\\
&+
\sin\theta_t
\left(
|\psi_{\perp,t}\rangle\langle\psi_t|
-
|\psi_t\rangle\langle\psi_{\perp,t}|
\right),
\label{eq:d2q5-ucol-t}
\end{align}
and its action on the instantaneous state is
\begin{equation}\label{eq:d2q5-post-t}
  U_{\mathrm{col},t}|\psi_t\rangle
  =
  \cos\theta_t|\psi_t\rangle
  +
  \sin\theta_t|\psi_{\perp,t}\rangle .
\end{equation}
Therefore, the angular distance from the equilibrium direction
changes from $\phi_t$ to
\begin{equation}\label{eq:d2q5-angle-relax}
  \phi_t-\theta_t
  =
  \left(1-\frac{1}{\tau}\right)\phi_t,
\end{equation}
which is consistent with the relaxation mechanism of the rotation
collision introduced in \cref{sec:rotation}.

The complete evolution of D2Q5 at time $t$ consists of the
rotation collision followed by the streaming operation,
\begin{equation}\label{eq:d2q5-step-t}
  U_{\mathrm{step},t}
  =
  U_{\mathrm{stream}}U_{\mathrm{col},t}.
\end{equation}
Following the sequential composition introduced in
\cref{sec:concat}, the evolution over $T$ time steps can
be written as
\begin{equation}\label{eq:d2q5-sequential}
  |\Psi(T)\rangle
  =
  U_{\mathrm{step},T-1}
  U_{\mathrm{step},T-2}
  \cdots
  U_{\mathrm{step},1}
  U_{\mathrm{step},0}
  |\Psi(0)\rangle ,
\end{equation}
or equivalently,
\begin{equation}\label{eq:d2q5-sequential-expanded}
  |\Psi(T)\rangle
  =
  \left[
  \prod_{t=T-1}^{0}
  U_{\mathrm{stream}}U_{\mathrm{col},t}
  \right]
  |\Psi(0)\rangle .
\end{equation}
It should be emphasized that the fixed equilibrium direction does
not imply that the complete rotation collision operator is time-independent. In general, the instantaneous state $|\psi_t\rangle$,
the angle $\phi_t$, the rotation angle $\theta_t$, and the associated
rotation subspace vary during the evolution. Thus,
$U_{\mathrm{col},t}$ may be different at successive time steps.

Nevertheless, the D2Q5 model provides an important circuit-level
simplification. Since the equilibrium direction
$|\hat e_{\mathrm{eq}}\rangle$ is fixed, the state-preparation
circuit associated with the equilibrium state and its inverse can
be compiled once and reused in every collision block. Furthermore,
for a fixed lattice and streaming implementation,
$U_{\mathrm{stream}}$ is also reused throughout the evolution.
Thus, only the state-dependent part of the rotation collision needs
to be updated as the distribution evolves.

Once the state-dependent parameters of the individual collision
blocks are specified, the resulting collision--streaming
blocks of D2Q5 model can be connected sequentially into a single unitary circuit, and no projective measurement is required solely for the composition
of adjacent unitary blocks. The coherent determination of the
state-dependent collision parameters from an intermediate quantum
state is, however, a separate algorithmic problem, and is not
addressed in the present construction.

\subsection{State-Dependent Rotation Collision for the D2Q9 Model}
\label{sec:d2q9-state-dependent}

For the D2Q9 model, the equilibrium distribution function is given
by~\cite{qian1992lattice}
\begin{equation}\label{eq:d2q9-eq}
  f_i^{\mathrm{eq}}
  =
  \rho w_i\!\left(
  1
  + \frac{\boldsymbol{e}_i\cdot\boldsymbol{u}}{c_s^2}
  + \frac{(\boldsymbol{e}_i\cdot\boldsymbol{u})^2}{2c_s^4}
  - \frac{|\boldsymbol{u}|^2}{2c_s^2}
  \right),
\end{equation}
where
$\boldsymbol{u}=\sum_j\boldsymbol{e}_j f_j/\rho$
is determined from the first-order moment of distribution function. Unlike the D2Q5 model,
the equilibrium distribution function depends nonlinearly on the current
state, and the normalized equilibrium state
$|\psi_{\mathrm{eq},t}\rangle$ varies during the evolution.
Consequently, the rotation angle and the collision operator
$U_{\mathrm{col},t}$ must be updated at each time step.

Although the evolution of D2Q9 model can still be written as a sequential
composition of collision--streaming blocks, the equilibrium-state
preparation circuit cannot be precompiled and reused as that in D2Q5 model.
In the present implementation, the distribution function is recovered by
measurement at each time step, after which $\boldsymbol{u}$,
$f_i^{\mathrm{eq}}$, and the corresponding rotation collision
operator are updated before the next step.

\subsection{Structure of the Sequential Collision Circuit}

According to \cref{sec:concat}, the $T$-step evolution of the
D2Q5 model is represented by the ordered sequence
\begin{equation}
  U_{\mathrm{step},T-1}\cdots U_{\mathrm{step},1}
  U_{\mathrm{step},0},
  \qquad
  U_{\mathrm{step},t}
  =
  U_{\mathrm{stream}}U_{\mathrm{col},t},
\end{equation}
where the streaming operator is fixed, while the rotation
collision operator generally depends on the instantaneous state.

The gate-level implementation of the collision operator is based on
the Shende--Bullock--Markov state preparation scheme (SBM)
\cite{shende2005synthesis} and phase reflection operators, with the
detailed SBM construction provided in
Appendix A. A
phase reflection about a state $|\phi\rangle$ is defined as
\begin{equation}\label{eq:reflection}
  S_\phi(\alpha)
  =
  I+(e^{\mathrm{i}\alpha}-1)|\phi\rangle\langle\phi|,
\end{equation}
which is unitary for any real $\alpha$. The collision operator at
time $t$ can be decomposed as
\begin{equation}\label{eq:ucol-refl}
  U_{\mathrm{col},t}
  =
  S_{\psi_t}(\gamma_t)
  S_{\mathrm{eq}}(\beta_t),
\end{equation}
where the phase parameters are determined from the instantaneous
angles $\phi_t$ and $\theta_t=\phi_t/\tau$. In particular,
\begin{equation}\label{eq:phase-params}
  \sin(\beta_t/2)
  =
  \frac{\sin\theta_t}{\sin 2\phi_t},
  \qquad
  \gamma_t
  =
  -\arg\!\left[
  e^{\mathrm{i}\alpha_{0,t}}
  \left(
  1+\cos^2\!\phi_t
  (e^{\mathrm{i}\beta_t}-1)
  \right)
  \right],
\end{equation}
where
$\alpha_{0,t}=-\pi/2-\beta_t/2$.
Thus, no numerical optimization is required once $\phi_t$ is
specified.

If $\mathrm{SBM}_\phi$ prepares $|\phi\rangle$ from $|0\rangle$,
the reflection can be implemented as
\begin{equation}\label{eq:sbm-expand}
  S_\phi(\alpha)
  =
  \mathrm{SBM}_\phi
  P_0(\alpha)
  \mathrm{SBM}_\phi^\dagger,
\end{equation}
where
\begin{equation}
  P_0(\alpha)
  =
  I+(e^{\mathrm{i}\alpha}-1)|0\rangle\langle0|
\end{equation}
is a multi-controlled phase operation. The collision circuit is
therefore implemented as
\begin{multline}\label{eq:circuit-expanded}
  |0\rangle
  \to \mathrm{SBM}_{\psi_t}
  \to \mathrm{SBM}_{\mathrm{eq}}^\dagger
  \to P_0(\beta_t)
  \to \mathrm{SBM}_{\mathrm{eq}}
  \\
  \to \mathrm{SBM}_{\psi_t}^\dagger
  \to P_0(\gamma_t)
  \to \mathrm{SBM}_{\psi_t}
  \to |\psi_t'\rangle .
\end{multline}
For the D2Q5 model, the normalized equilibrium direction
$|\hat e_{\mathrm{eq}}\rangle$ is independent of the instantaneous
distribution function. Consequently,
$\mathrm{SBM}_{\mathrm{eq}}$ and
$\mathrm{SBM}_{\mathrm{eq}}^\dagger$ can be compiled once and
reused in every collision block. By contrast,
$\mathrm{SBM}_{\psi_t}$, the phase parameters
$\beta_t$ and $\gamma_t$ generally vary with the instantaneous
state.

Once the state-dependent parameters of the individual collision
blocks are specified, these unitary collision--streaming blocks
can be connected sequentially without projective measurement
solely for their composition. The coherent determination of these
state-dependent parameters during evolution is a separate
algorithmic problem.

\section{Numerical results}
\label{sec:experiments}

\subsection{D2Q9 Model: Taylor--Green Vortex}

The first benchmark we consider is the two-dimensional Taylor--Green vortex,
which admits an exact solution of the incompressible Navier--Stokes
equations defined on the periodic domain $[0,1]^2$,
\begin{equation}\label{eq:tgv}
  \begin{aligned}
    u_x(x,y,t) &= -U_0\cos(2\pi x)\sin(2\pi y)\,e^{-2\nu(2\pi)^2 t},\\
    u_y(x,y,t) &= \phantom{-}U_0\sin(2\pi x)\cos(2\pi y)\,e^{-2\nu(2\pi)^2 t}.
  \end{aligned}
\end{equation}
The exponential decay is governed by viscous dissipation and involves
fully nonlinear Navier--Stokes dynamics, which cannot be represented within the reduced D2Q5 velocity set, and the D2Q9 model should be adopted. The simulation parameters are $N=32$, $Re=100$, $U_0=0.1$, and
$\tau=0.596$. The initial condition of distribution function is constructed by its
equilibrium part with the exact velocity field
\cref{eq:tgv} at $t=0$, and the periodic boundary conditions are imposed in both
spatial directions. To evaluate the difference between the rotation collision operator and the
classical BGK operator, we consider the vorticity
$\omega = \partial_x u_y - \partial_y u_x$, approximated using a
second-order central difference scheme. The comparison is performed at the half-life time $t_{1/2}$,
defined as the time at which the amplitude of the velocity field
decays to half of its initial value,
i.e.,
$U(t_{1/2})/U_0=1/2$.

\begin{figure}[htbp]
  \centering
  \label{fig:a}\includegraphics[width=\textwidth]{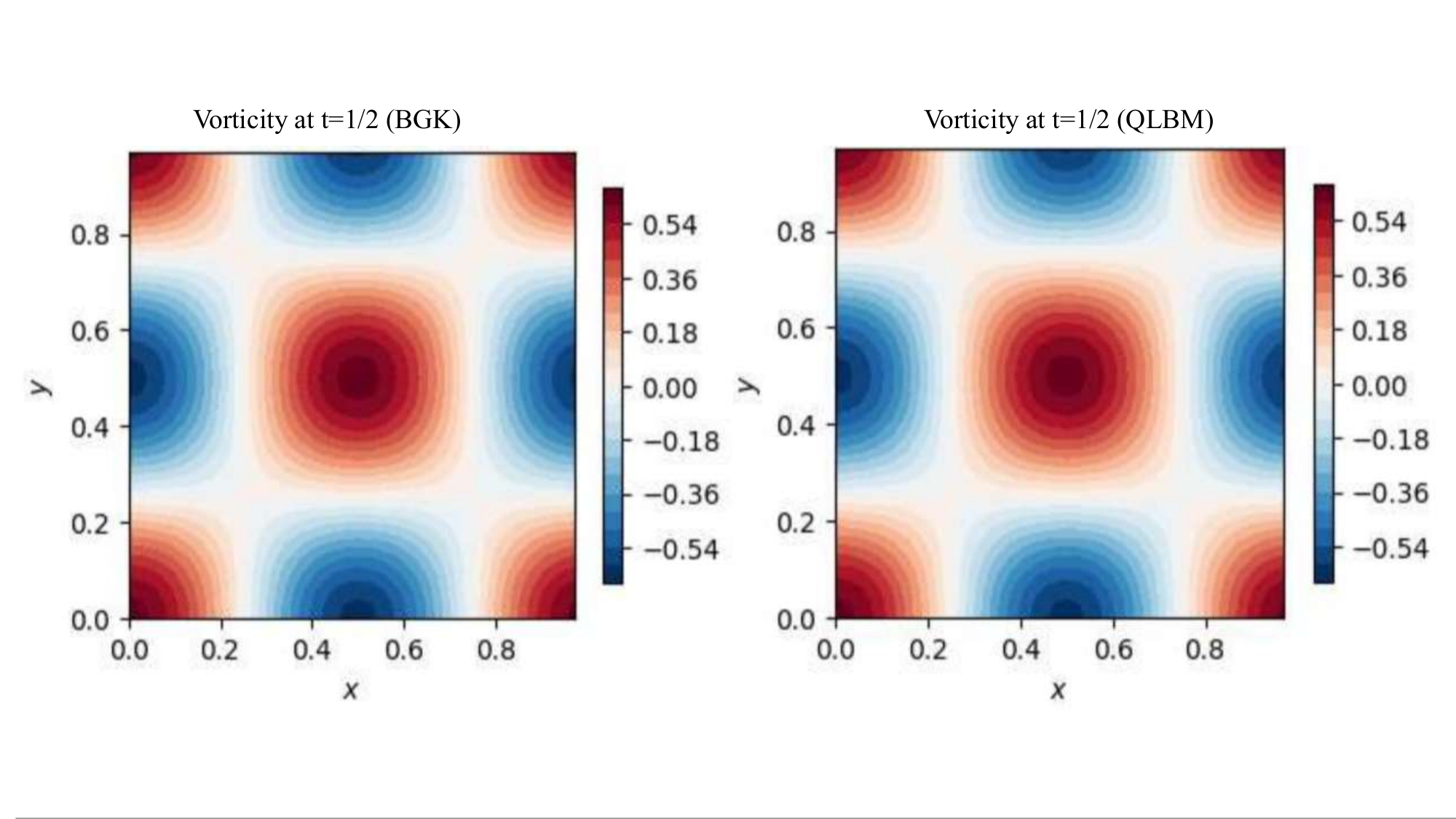}
  \caption{Vorticity contours for the Taylor--Green vortex at
  $Re=100$, $N=32$, $U_0=0.1$, $\tau=0.596$, evaluated at the
  half-life time $t_{1/2}$ of the velocity decay. Left: vorticity
  obtained by the classical BGK collision operator. Right:
  vorticity obtained by the rotation collision operator}
  \label{fig:fig1}
\end{figure}

\cref{fig:fig1} shows the vorticity fields obtained from both collision operators.
From this figure, one can observe that the two vorticity distributions exhibit nearly identical spatial patterns,
including the locations and signs of the vortices, indicating that
the rotation collision operator preserves the essential dissipative dynamics
of the classical BGK model. We also note that there are some minor differences in the
vorticity magnitude, which arise from the nonlinear mapping between the
amplitude evolution in the quantum representation and the reconstruction of distribution function. These results demonstrate that the proposed
unitary collision operator provides a consistent approximation to the
classical BGK relaxation for the Taylor--Green vortex.

\subsection{D2Q9 Model: Poiseuille Flow}
The second benchmark is the two-dimensional plane Poiseuille flow driven by an uniform body force, and the cases with three different Reynolds numbers, i.e., $Re = 10$, $40$, and $100$, are used to test the D2Q9 model. In the present QLBM, the collision step is implemented as a unitary rotation operator on the quantum state, as described in \crefname{subsection}{Section}{Sections}\cref{sec:rotation}. The streaming step and the boundary conditions can be handled within a fully quantum framework via the singular value decomposition approach \cite{kumar2025quantum}, which embeds non-unitary operators into unitary dilations. In the present work, however, both the streaming step and the half-way bounce-back scheme are treated classically , and the forcing scheme developed by Guo et al. \cite{guo2002discrete} is adopted to incorporate the effect of body force. 

\begin{figure}[htbp]
  \centering
  \label{fig:a}\includegraphics[width=\textwidth]{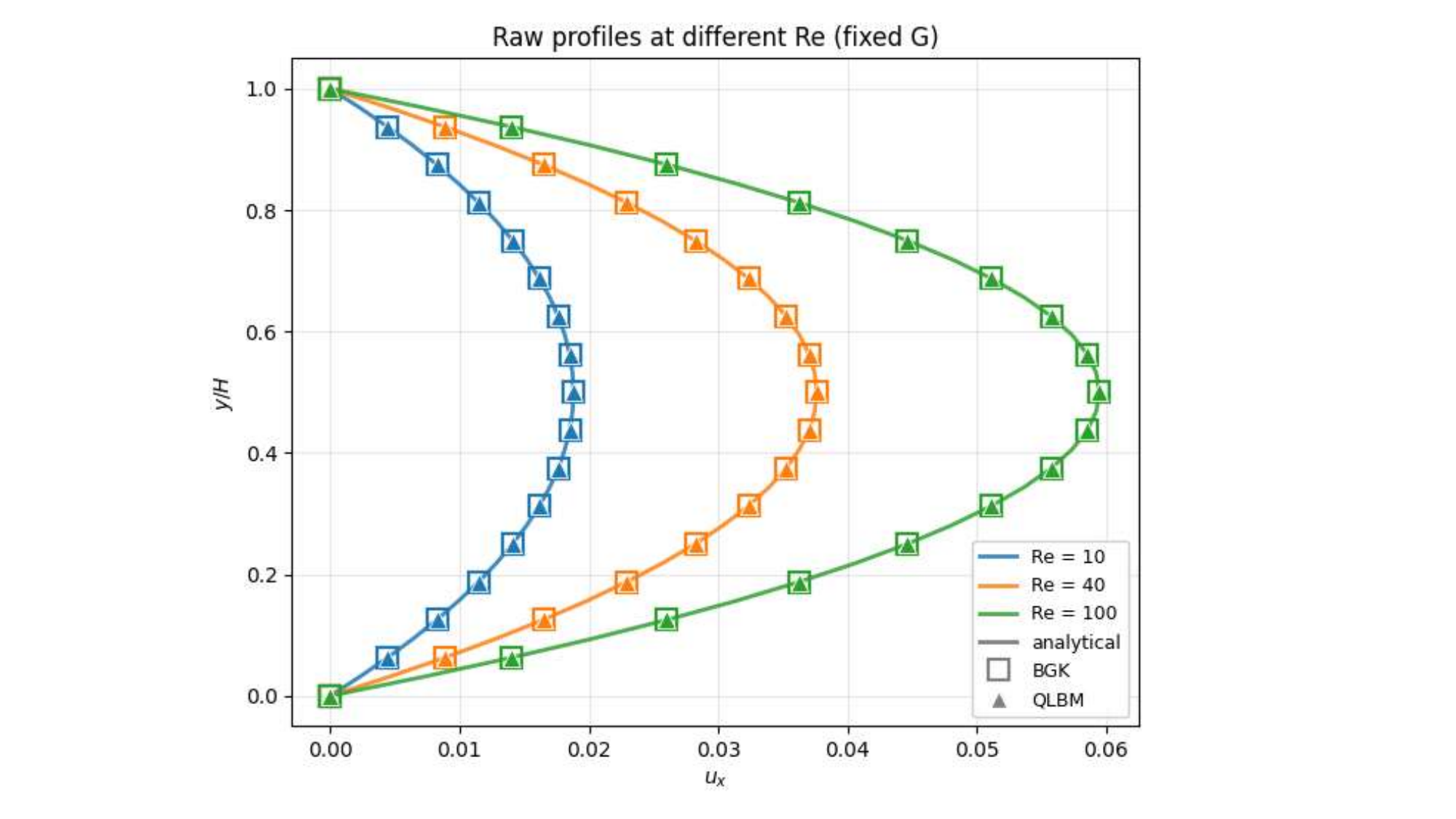}
  \caption{Velocity profiles of three cases with $Re=10,40,100$.
The results of lattice BGK method and QLBM agree with the analytical solution.}
  \label{fig:fig2}
\end{figure}

The driving force is fixed for all three cases, and the kinematic viscosity is varied to achieve the different Reynolds numbers.
\Cref{fig:fig2} presents the velocity profiles $u_x(y)$ at $Re = 10$, $40$, and $100$, and the numerical results are in good agreement with the analytical
solution
\begin{equation}
    u_x(y) = \frac{G}{2\nu}\,y\!\left(H - y\right),
    \label{eq:poiseuille}
\end{equation}
and also, the classical lattice BGK method and QLBM yield almost the same results.

\subsection{Error Scaling with Time Step and Relaxation Time}
\label{sec:error-scaling}

To quantify the discrepancy between the rotation collision and the
classical BGK collision, we perform two parameter sweeps using the
two-dimensional Taylor--Green vortex on a uniform grid $N\times N$
with $N=32$. In the first test, the time step $\delta t$ is varied
with a fixed $\tau=0.7$, while in the second test, $\tau$ is varied at
a fixed $U_0=0.1$. The lattice BGK method and QLBM use same
initial conditions, streaming procedures, and boundary conditions.

To evaluate the discrepancy, the following relative velocity errors are adopted
\begin{subequations}
\label{eq:relative-error-definitions}
\begin{align}
E_{\mathrm{BGK,ex}}^{u}
&=
\frac{
\|\boldsymbol{u}^{\mathrm{BGK}}
-\boldsymbol{u}^{\mathrm{ex}}\|_{2,h}
}{
\|\boldsymbol{u}^{\mathrm{ex}}\|_{2,h}
},
\\
E_{\mathrm{Q,ex}}^{u}
&=
\frac{
\|\boldsymbol{u}^{\mathrm{Q}}
-\boldsymbol{u}^{\mathrm{ex}}\|_{2,h}
}{
\|\boldsymbol{u}^{\mathrm{ex}}\|_{2,h}
},
\\
E_{\mathrm{Q,BGK}}^{u}
&=
\frac{
\|\boldsymbol{u}^{\mathrm{Q}}
-\boldsymbol{u}^{\mathrm{BGK}}\|_{2,h}
}{
\|\boldsymbol{u}^{\mathrm{BGK}}\|_{2,h}
},
\end{align}
\end{subequations}
where $\|\cdot\|_{2,h}$ denotes the discrete $L_2$ norm.

\begin{figure}[htbp]
  \centering
  \includegraphics[width=\textwidth]{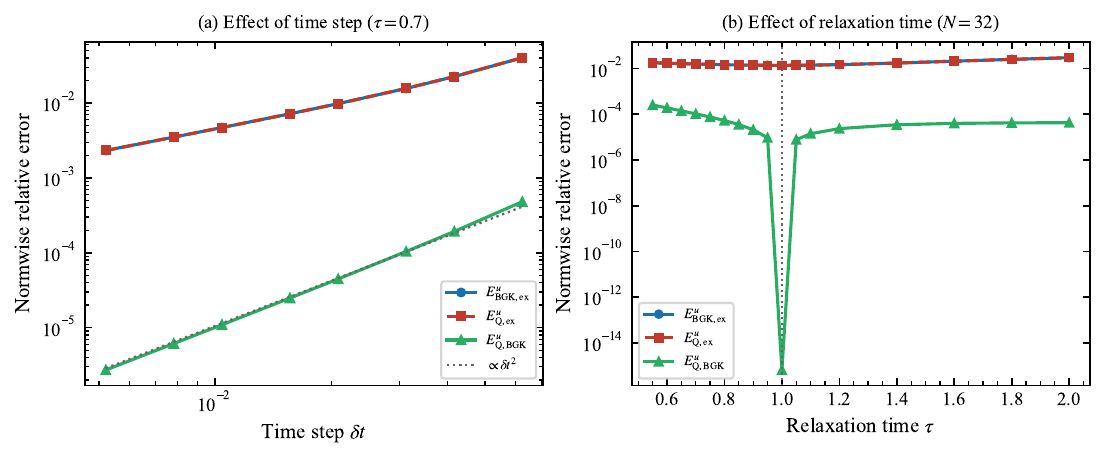}
  \caption{
  Relative errors of velocity based on the lattice BGK method and QLBM.
  (a) Effect of time step ($\tau=0.7$), the dotted line
  denotes the reference scaling $\alpha\delta t^2$.
  (b) Effect of relaxation time ($N=32$ and $U_0=0.1$); the two
  collision operators coincide at $\tau=1$ and discrepancy between them is up to floating-point
  round-off.
  }
  \label{fig:error-dt}
\end{figure}

Figure~\ref{fig:error-dt}(a) shows the realtive errors at different time steps, and the slope of dotted line is 2. From this figure, one can see that the errors of the lattice BGK  method is almost the same as those of QLBM, and the discrepancy
$E_{\mathrm{Q,BGK}}^{u}$ between the rotation collision and the classical BGK collision follows the scaling
\begin{equation}
  E_{\mathrm{Q,BGK}}^{u}=O(\delta t^2),
\end{equation}
which is consistent with the theoretical analysis in \cref{sec:rotation}.

Figure~\ref{fig:error-dt}(b) presents the effect of the relaxation
time on the numerical results. It is also found that these two methods almost give the same results, while $E_{\mathrm{Q,BGK}}^{u}$ reaches the
floating-point round-off level at $\tau=1$. As stated previously, when $\tau=1$,
$\theta=\phi$ so that both collision operators map the distribution function
exactly to its equilibrium part. However, if $\tau\neq 1$, the discrepancy
increases due to the higher-order difference between the two
collision formulations.

Overall, the rotation collision preserves the accuracy of lattice
BGK method, and exhibits the predicted second-order discrepancy with
respect to $\delta t$.

\section{Discussion}
\label{sec:discussion}

\paragraph{Circuit reuse and multi-step composition}
For the D2Q5 model, the linear form of the equilibrium distribution function can
render the normalized equilibrium state independent of the
instantaneous distribution, and in this case the equilibrium-state
preparation circuit can be compiled once and reused throughout the
evolution. The complete collision operator, however, generally depends
on the instantaneous quantum state through the corresponding rotation
parameters.

Once the state-dependent parameters of the individual collision blocks
are specified, the unitary collision--streaming blocks can be composed
sequentially without projective measurement solely for their
composition, thereby preserving quantum coherence during the
corresponding unitary evolution.

\paragraph{State-dependent equilibrium distribution function of the D2Q9 model}
In contrast, the equilibrium distribution function of D2Q9 model depends nonlinearly on
the local macroscopic velocity, which causes both the equilibrium
state and collision operator to vary with the current state. In the present hybrid
implementation, the macroscopic variables are reconstructed from the
current distribution function and used classically to update the equilibrium
state and collision parameters at each time step. A fully coherent
determination of these state-dependent quantities would require
additional quantum subroutines.

\paragraph{Comparison with existing approaches and limitations}
Compared with existing quantum collision schemes, the present
rotation-based collision operator requires no ancilla qubits or
post-selection \cite{mezzacapo2015quantum}, and does not need to remove the
collision step \cite{todorova2020quantum},
and additionally, the approximation of the non-equilibrium distribution function based on the Chapman-Enskog expansion~\cite{zeng2025quantum} can also be avoided. Moreover, the
difference between the rotation collsion and BGK collision is of the order
$O(\delta t^2)$.

We would also like to point out that the present QLBM still has several
limitations. First, the complexity analysis does not include the cost
of state preparation and terminal measurement, which
scales as $O(N)$ and may partially offset the quantum advantage.
Second, the streaming step and boundary conditions are treated
classically. How to realize the quantum implementation of them remains an open
problem, which may be potentially addressable via the singular value decomposition
framework \cite{kumar2025quantum}. Finally, all results are obtained
via classical simulations, and the impact of hardware noise on real
quantum devices also needs to be investigated.
\section{Conclusion}
\label{sec:conclusion}

In this work, we developed a quantum lattice Boltzmann framework based on a
rotation collision operator, providing a unitary formulation of the classical
BGK relaxation step. The operator is implemented as a rotation in the
two-dimensional subspace spanned by the current state and its equilibrium
counterpart, and is unitary by construction. It is also shown that the difference between the rotation collision and the classical BGK
operator is of the order $O(\delta t^2)$.

A key theoretical result of this work is the reusable equilibrium-state
structure of the D2Q5 model. Owing to the linear dependence of its equilibrium
distribution on the conservative scalar $\phi$, the normalized equilibrium state
is independent of the distribution function, allowing the corresponding
equilibrium-state preparation circuit to be compiled once and reused throughout
the evolution. The complete collision operator, however, generally depends on
the instantaneous state. Accordingly, the $T$-step evolution is represented by
the ordered composition
$|\Psi(T)\rangle =
U_{\mathrm{step},T-1}\cdots U_{\mathrm{step},0}|\Psi(0)\rangle$ with  $U_{\mathrm{step},t}
=U_{\mathrm{stream}}U_{\mathrm{col},t}$.
In contrast, the equilibrium distribution function of D2Q9 model depends on the 
macroscopic velocity, so that both the
equilibrium state preparation and collision parameters generally vary during the evolution.

Numerical results of the Taylor--Green vortex and Poiseuille flow
demonstrate that there is a good agreement between the proposed QLBM and the classical lattice
BGK method, and the difference between them is of the order $O(\delta t^2)$. The circuit implementation of D2Q5 model,
constructed via the Shende--Bullock--Markov state preparation scheme and
phase reflection operator, is provided in the appendix together with
explicit gate-count comparison.

Overall, the present work establishes a unitary and gate-level realizable
formulation of lattice Boltzmann method, and identifies the reusable
equilibrium-state circuit structure of the D2Q5 model as an advantageous
feature for quantum implementation.
\appendix

\section{Shende--Bullock--Markov State Preparation Circuit}
\label{sec:appendix-sbm}

The Shende--Bullock--Markov (SBM) scheme
\cite{shende2005synthesis} provides a systematic construction for
quantum-state preparation. In the present work, the target amplitudes
are real and non-negative because they are defined by
$a_i=\sqrt{f_i/\rho}$, thus only the real-amplitude part of the
SBM construction is required. The resulting state-preparation circuit
provides the basic primitive used to implement the phase reflections
in the D2Q5 collision circuit.

\subsection{Problem statement}

Given a normalized real state
\begin{equation}
  |\phi\rangle
  =
  \sum_{k=0}^{2^n-1}\phi_k|k\rangle,
  \qquad
  \phi_k\geq0,
  \qquad
  \sum_k\phi_k^2=1,
\end{equation}
the objective is to construct a unitary
$\mathrm{SBM}_\phi$ satisfying
\begin{equation}
  \mathrm{SBM}_\phi|0\rangle^{\otimes n}
  =
  |\phi\rangle .
\end{equation}

\subsection{Recursive construction}

The circuit is constructed recursively from the most significant
qubit (level $\ell=0$) to the least significant qubit
(level $\ell=n-1$). At the level $\ell$, the amplitudes are divided into
$2^\ell$ blocks, each of them is further separated into left and
right halves,
$\boldsymbol{\phi}^{(g)}_L$ and
$\boldsymbol{\phi}^{(g)}_R$.
The corresponding rotation angle is
\begin{equation}\label{eq:sbm-angle}
  \vartheta_{\ell,g}
  =
  2\arctan\!\left(
  \frac{\|\boldsymbol{\phi}^{(g)}_R\|}
       {\|\boldsymbol{\phi}^{(g)}_L\|}
  \right),
\end{equation}
where the zero-norm branches are omitted. The gate at level $\ell$ is an
$R_y(\vartheta_{\ell,g})$ rotation on qubit
$q=n-1-\ell$, controlled by the preceding $\ell$ qubits according
to the binary pattern associated with block $g$. Thus, $\ell=0$
corresponds to an uncontrolled $R_y$ rotation, whereas higher levels
involve controlled rotations.

\subsection{Three-qubit case (D2Q5 model)}

For the D2Q5 model, three qubits are used to embed the five discrete
velocity populations into the eight-dimensional state
\begin{equation}
  |\phi\rangle
  =
  \sum_{i=0}^{4}\phi_i|i\rangle ,
\end{equation}
where the remaining amplitudes are set to be zero.

At the level $\ell=0$, one $R_y$ rotation is applied to $q_2$.
At the level $\ell=1$, two controlled $R_y$ rotations act on $q_1$,
conditioned on $q_2=0$ and $q_2=1$, respectively.
At the level $\ell=2$, the non-zero amplitude pairs
$(\phi_0,\phi_1)$ and $(\phi_2,\phi_3)$ give two nontrivial
doubly controlled rotations on $q_0$. The remaining padded branches
produce zero rotation angles and are omitted.

For the equilibrium state in D2Q5 model,
\begin{equation}
  |\hat e_{\mathrm{eq}}\rangle
  \propto
  \sum_i\sqrt{\lambda_i}|i\rangle ,
\end{equation}
the amplitudes are independent of the instantaneous distribution functions.
Consequently, $\mathrm{SBM}_{\mathrm{eq}}$ and its inverse can be
compiled once and reused in every collision block.

By contrast, $\mathrm{SBM}_{\psi_t}$ depends on the instantaneous
state. Once the amplitudes defining $|\psi_t\rangle$ are specified,
the rotation angles can be obtained directly from
\cref{eq:sbm-angle}. The determination of these amplitudes during a
fully coherent multistep evolution is separate from the SBM
state-preparation procedure.

\begin{figure}[htbp]
  \centering
  \scalebox{0.82}{%
  \begin{quantikz}[row sep=0.7cm, column sep=0.4cm]
    \lstick{$q_2$} &
      \gate{R_y(\vartheta_{0,0})} &
      \octrl{1} &
      \ctrl{1} &
      \octrl{1} &
      \octrl{1} &
      \qw \\
    \lstick{$q_1$} &
      \qw &
      \gate{R_y(\vartheta_{1,0})} &
      \gate{R_y(\vartheta_{1,1})} &
      \octrl{1} &
      \ctrl{1} &
      \qw \\
    \lstick{$q_0$} &
      \qw & \qw & \qw &
      \gate{R_y(\vartheta_{2,0})} &
      \gate{R_y(\vartheta_{2,1})} &
      \qw
  \end{quantikz}}
  \caption{
  Three-qubit SBM circuit for the amplitude encoding of D2Q5 model.
  Qubits are ordered from $q_2$ (MSB) to $q_0$ (LSB).
  Open and filled controls denote conditions on $|0\rangle$ and
  $|1\rangle$, respectively. Gates associated with zero-padded
  branches are omitted.
  }
  \label{fig:sbm-circuit}
\end{figure}
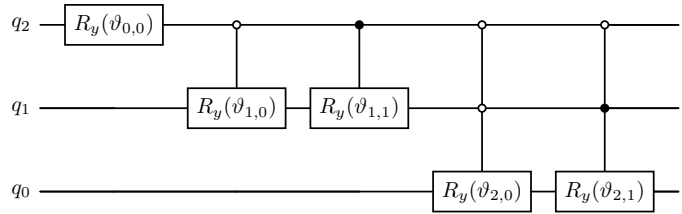
\section{Collision Circuit Construction of the D2Q5 Model}
\label{sec:appendix-collision}

The collision circuit of D2Q5 model is implemented by using generalized phase
reflections. At time $t$, the prescribed rotation angle is
$\theta_t=\phi_t/\tau$, where
$\phi_t=\arccos\langle\psi_t|\hat e_{\mathrm{eq}}\rangle$.

\subsection{Phase-reflection realization}

For a normalized state $|\phi\rangle$, the generalized phase
reflection is defined as
\begin{equation}\label{eq:reflection-def}
  S_\phi(\alpha)
  =
  I+
  \left(e^{\mathrm{i}\alpha}-1\right)
  |\phi\rangle\langle\phi|.
\end{equation}
Let $\mathrm{SBM}_\phi$ satisfy
$\mathrm{SBM}_\phi|0\rangle^{\otimes n}=|\phi\rangle$,
then we have
\begin{equation}\label{eq:reflection-expand}
  S_\phi(\alpha)
  =
  \mathrm{SBM}_\phi
  P_0(\alpha)
  \mathrm{SBM}_\phi^\dagger,
\end{equation}
where
\begin{equation}
  P_0(\alpha)
  =
  I+
  \left(e^{\mathrm{i}\alpha}-1\right)
  |0\rangle\langle0|.
\end{equation}
where the phase $e^{\mathrm{i}\alpha}$ has been applied to
$|0\rangle^{\otimes n}$.

The collision operator can be decomposed as
\begin{equation}\label{eq:ucol-two-refl}
  U_{\mathrm{col},t}
  =
  S_{\psi_t}(\gamma_t)
  S_{\mathrm{eq}}(\beta_t).
\end{equation}
The phase parameters are chosen so that the two reflections generate
the desired rotation in
$\operatorname{span}\{|\psi_t\rangle,|\hat e_{\mathrm{eq}}\rangle\}$:
\begin{equation}\label{eq:phase-params}
  \sin\!\left(\frac{\beta_t}{2}\right)
  =
  \frac{\sin\theta_t}{\sin 2\phi_t},
  \qquad
  \gamma_t
  =
  -\arg\!\left[
    e^{\mathrm{i}\alpha_{0,t}}
    \left(
      1+
      \cos^2\!\phi_t
      \left(e^{\mathrm{i}\beta_t}-1\right)
    \right)
  \right],
\end{equation}
where
\begin{equation}
  \alpha_{0,t}
  =
  -\frac{\pi}{2}
  -
  \frac{\beta_t}{2}.
\end{equation}
Thus, once $\phi_t$ is specified, $\beta_t$ and $\gamma_t$ can be
obtained directly without numerical optimization.

With the help of \cref{eq:reflection-expand}, the single-step collision circuit
takes the form
\begin{multline}\label{eq:circuit-expanded}
  |\psi_t\rangle
  \to
  \mathrm{SBM}_{\mathrm{eq}}^\dagger
  \to P_0(\beta_t)
  \to \mathrm{SBM}_{\mathrm{eq}}
  \\
  \to
  \mathrm{SBM}_{\psi_t}^\dagger
  \to P_0(\gamma_t)
  \to \mathrm{SBM}_{\psi_t}
  \to |\psi_t'\rangle .
\end{multline}
For the D2Q5 model, $|\hat e_{\mathrm{eq}}\rangle$ is fixed, thus
$\mathrm{SBM}_{\mathrm{eq}}$ and its inverse can be compiled once
and reused. By contrast, $\mathrm{SBM}_{\psi_t}$, $\beta_t$, and
$\gamma_t$ generally depend on the instantaneous state.

\subsection{Sequential multi-step circuit}

A complete time step is
\begin{equation}\label{eq:appendix-step-t}
  U_{\mathrm{step},t}
  =
  U_{\mathrm{stream}}
  U_{\mathrm{col},t},
\end{equation}
and the $T$-step evolution is
\begin{equation}\label{eq:appendix-d2q5-multistep}
  |\Psi(T)\rangle
  =
  U_{\mathrm{step},T-1}
  \cdots
  U_{\mathrm{step},1}
  U_{\mathrm{step},0}
  |\Psi(0)\rangle .
\end{equation}
The collision blocks are generally state-dependent, while the
equilibrium-state preparation circuit and the streaming operator are
reused throughout the evolution. Once the parameters of the
individual collision blocks are specified, these unitary blocks can
be connected sequentially without projective measurement solely for
their composition.

\begin{figure}[htbp]
  \centering
  \scalebox{0.88}{%
  \begin{quantikz}[row sep=0.65cm, column sep=0.45cm]
    \lstick{$|\Psi(0)\rangle$} &
    \gate[3]{U_{\mathrm{col},0}} &
    \gate[3]{U_{\mathrm{stream}}} &
    \gate[3]{U_{\mathrm{col},1}} &
    \gate[3]{U_{\mathrm{stream}}} &
    \qw &
    \cdots &
    \qw &
    \gate[3]{U_{\mathrm{col},T-1}} &
    \gate[3]{U_{\mathrm{stream}}} &
    \meter{} \\
    \qw & \qw & \qw & \qw & \qw & \qw & \qw & \qw & \qw & \qw & \qw \\
    \qw & \qw & \qw & \qw & \qw & \qw & \qw & \qw & \qw & \qw & \qw
  \end{quantikz}}
  \caption{
  Sequential $T$-step circuit of D2Q5 model.
  Each step consists of a state-dependent collision
  $U_{\mathrm{col},t}$ followed by the fixed streaming operator.
  }
  \label{fig:concat-circuit}
\end{figure}
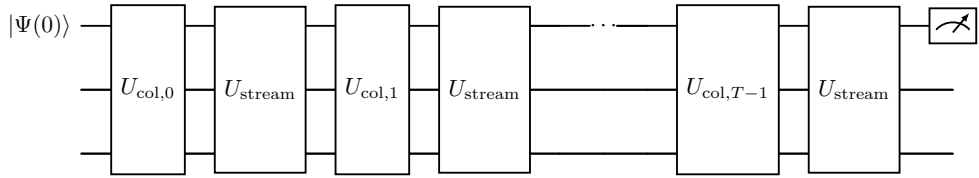
%--------------------------------------------------------------
\begin{table}[htbp]
  \centering
  \caption{The collision circuit resources after transpilation to the
$\{U_3,CX\}$ basis with optimization level 0 (state preparation excluded).}
  \label{tab:d2q9-resources}
  \begin{tabular}{lrr}
    \hline
    Metric            & UnitaryGate & DiagonalGate \\
    \hline
    Qubits            &  4  &  5  \\
    $CX$ gates        & 95  & 196 \\
    $U_3$ gates       & 180 & 286 \\
    Total gates       & 275 & 482 \\
    Circuit depth     & 189 & 328 \\
    \hline
  \end{tabular}
\end{table}
\section{The Collision Circuit Resource of D2Q9 model}
\label{sec:appendix-d2q9}

Two quantum implementations of the rotation collision operator in D2Q9 model
are compared. Method A: a \texttt{UnitaryGate} to apply the full
$16\times16$ unitary matrix $U_{\mathrm{col}}$ directly on 4 qubits;
Method B: $U_{\mathrm{col}}$ \cite{welch2014efficient}, encode the state in the
eigenbasis, and applly the eigenphases.

Both circuits were transpiled to the $\{U_3, CX\}$ basis at
optimization level~0 by using Qiskit \cite{fingerhuth2018open}.
\Cref{tab:d2q9-resources} reports the gate counts and circuit depth
for the collision subcircuit only, with state preparation excluded.

The UnitaryGate approach requires approximately half number of
$CX$ gates compared with the DiagonalGate approach
(ratio $196/95 \approx 2.06$), while using one fewer qubit.
Both methods require a projective measurement at each time step to
extract the macroscopic velocity $\boldsymbol{u}^{(t)}$ and
reconstruct $U_{\mathrm{col}}^{(t+1)}$, consistent with the
non-concatenability of the D2Q9 model in
\cref{sec:d2q9-state-dependent}.

\bibliographystyle{siamplain}
\bibliography{references}
\end{document}